\documentstyle[12pt,eepic,epic]{article}
\newlength{\minitwocolumn}
\def\nb{\mbox{\boldmath $n$}}
\def\pb{\mbox{\boldmath $p$}}
\def\l{\ell}
\def\c{{\cal C}}
\def\pv{\hat{P}_V}
\def\fv{\hat{f}_V}
\def\q{\hat{q}}
\def\s{\hat{s}}
\def\u{\hat{u}} 
\def\ml{\hat{m}_\ell}
\def\ms{\hat{m}_s}
\def\md{\hat{m}_d}
\def\mc{\hat{m}_c}
\def\muq{\hat{m}_u}
\def\mq{\hat{m}_q}
\def\mv{\hat{m}_V}
\def\mui{\hat{m}_{u^i}}

\title{Semileptonic $B-$decay as a test of CKM unitarity}
\author{\vspace{5mm}\\
        {\bf L. T. Handoko}\thanks{
        E-mail address : lthandoko@bigfoot.com} \\
        \vspace{2mm}\\
        Laboratory for Theoretical Physics and Mathematics \\
        Indonesian Institute of Sciences \\
        Kom. Puspitek Serpong P3FT--LIPI, Tangerang 15310, Indonesia\\
        and \\
        Department of Physics, Hiroshima University \\
        1-3-1 Kagamiyama, Higashi Hiroshima 739-0046, Japan}
\date{}

\begin{document}

\setlength{\baselineskip}{24pt}

\maketitle
\begin{picture}(0,0)
       \put(310,315){LFTMLIPI-139701}
       \put(310,300){HUPD-9714}
       \put(310,285){August 1997}
\end{picture}

\setlength{\baselineskip}{8mm}
\thispagestyle{empty}

\begin{abstract}
	I point out that $B \rightarrow X_q \, \l^+ \, \l^-$ 
	decays $(q = s, d)$ 
	are sensitive probes of possible violation of CKM unitarity.
	I compute the decay rates and asymmetries in a
	minimal extension of the 
	Standard Model containing an additional isosinglet charge 
	(-1/3) quark, which leads to a deviation from CKM
	unitarity. 
It is shown that even for small mixing ratios 
$\left| {z_{qb}}/{(V_{tq}^\ast V_{tb})} \right| \sim O({10}^{-2})$, 
the contribution of the tree-level $Z-$FCNC appearing in the
model should change the rates and asymmetries significantly. Especially	 
the CP asymmetry, ${\cal A}_{\rm CP} (B \rightarrow X_s \, \l^+ \, \l^-)$, 
can be enhanced to be few percents, while in the 
standard model the size is less than $O({10}^{-3})$. 
On the other hand, 
${\cal A}_{\rm CP} (B \rightarrow X_d \, \l^+ \, \l^-)$ is
not altered so much. 
Constraints for the mixing ratios are extracted 
from the experiments of $B \rightarrow X_s \,\gamma$ for 
$q = s$ and $B_d^0 - \bar{B}_d^0$ mixing for $q = d$
under a natural assumption that the couplings of the
tree-level $Zf\bar{f}$ are almost unity, i.e. $z_{\alpha \alpha} \sim 1$.
\end{abstract}

\vspace*{5mm}
PACS number(s) : 12.15.Hh; 12.60.Cn; 13.20.He; 13.25.Hw; 13.30.Ce

\clearpage

Beyond the experimentally measured $B \rightarrow X_s \, \gamma$, 
the study of another flavor-changing neutral current (FCNC) processes, 
$B \rightarrow X_q \, \l^+ \, \l^-$ with $q = s$ or $d$, may
also provide important test of the Standard Model (SM) as
well as open a window for physics beyond it.  

In the present letter, I point out that these decays are sensitive  
and has crucial dependence on the unitarity of the 
Cabibbo-Kobayashi-Maskawa (CKM) matrix.
For definiteness, I adopt a typical model that violates the unitarity,  
that is the SM containing an additional isosinglet charge (-1/3)
quark \cite{vlq}. Then the decay rates and asymmetries of the channels
will be analysed in the framework of the model. 
Phenomenologically, the unitarity violation in the new CKM 
matrix is not affecting the branching fraction $B \rightarrow X_s \, \gamma$
to any level of significance \cite{bqg}. However, I will show 
that $B \rightarrow X_q \, \l^+ \, \l^-$ even for $q = s$ can 
change significantly. Furthermore, there are also theoretically 
appealing motivations behind its consideration, i.e. 
$E_6$ GUT, some superstring-inspired model and a solution to
the strong CP problem \cite{tvlq}. 

Now I briefly describe the model. An extra down-type (charge
-1/3) quark, whose left and right handed components are 
both SU(2) singlets, is introduced. Then the particle
content for the quark sector is, 
 \[
	\left(
	\begin{array}{c}
		u_i \\
		d_i
	\end{array}
	\right)_L \: , \:
        {d_L}_4 \: , \: {u_R}_i \: , \: {d_R}_\alpha
\]
with generation indices $\alpha = 1,2,3,4$ and $i = 1,2,3$. 
Throughout this paper I use the following notations for
chirality, $L/R \equiv {(1 \mp \gamma_5)}/2$. Consequently, 
in the present model the gauge interactions in terms of the 
mass eigenstates become, 
\begin{eqnarray}
        {\cal L}_{W^{\pm}} & = & \frac{g}{\sqrt{2}} \, 
                V_{i \alpha} \bar{u}_i \, \gamma^\mu \, L \,
                d_\alpha \, W_{\mu}^+ + {\rm h.c.} , \\
	\label{eq:w}
	{\cal L}_Z & = & \frac{g}{2 \cos \theta_W} \, 
		\bar{d}_{\alpha} \, \gamma^{\mu} \left[
		\left( \frac{2}{3} \sin^2 \theta_W \, \delta_{\alpha\beta} -
                z_{\alpha \beta} \right) L + 
			\frac{2}{3} \sin^2 \theta_W \, \delta_{\alpha\beta} \, 
                R \right] d_\beta \, Z_{\mu} \; , 
	\label{eq:z}
\end{eqnarray}
where
\begin{equation}
	z_{\alpha \beta} \equiv \sum_{i=1}^{3} {U^d}_{\alpha i}
                {U^d}_{i \beta}^{\ast} 
	= \delta_{\alpha\beta} - {U^d}_{\alpha 4} {U^d}_{4 \beta}^{\ast} \; , 
	\label{eq:zab}
\end{equation}
and the new CKM matrix is a $3 \times 4$ one, that is 
\begin{equation}
	V_{i \alpha} = \sum_{j = 1}^{3} 
		{U^u}_{ij} \, {{U^d}_{j \alpha}}^\ast \; . 
	\label{eq:ckm}
\end{equation}
Here $U^{u,d}$ are unitary matrices that relate
the weak and mass eigenstates as below,  
\begin{equation}
	\left(
	\begin{array}{c}
		u_i \\
		d_\alpha 
	\end{array}
	\right)_{\rm mass} = 
	\left(
	\begin{array}{cc}
		{U^u}_{ij} 	& 0 \\
		0		& {U^d}_{\alpha \beta}
	\end{array}
	\right) \; \left( 
	\begin{array}{c}
		u_j \\
		d_\beta 
	\end{array}
	\right)_{\rm weak} \; .
	\label{eq:weakmass} 
\end{equation}
Indeed, using the unitarity of $U^{u,d}$, the CKM unitarity 
violation in the present interest can be expressed as follows,
\begin{equation}
	V_{uq}^\ast \, V_{ub} + V_{cq}^\ast \, V_{cb} + 
	V_{tq}^\ast \, V_{tb} = z_{qb} \; .
	\label{eq:vckm}
\end{equation}
From Eq. (\ref{eq:z}), it is clear that tree-level FCNC is appearing 
in the model. Of course other tree-level FCNC also 
exist in the neutral Higgs sector, but it should be suppressed 
by a factor $1/{M_W}$. So I ignore the contribution here. 
Remark that the photon interaction is not altered at all. 

In most of models, including the SM and the present model,  
$b \rightarrow q \, \l^+ \, \l^-$ decays can be expressed 
by three effective Wilson coefficients and are governed 
by the following effective Hamiltonian \cite{bqll}, 
\begin{eqnarray}
        {\cal H}_{\rm eff} & = & \frac{G_F \, \alpha}{\sqrt{2} \, \pi} \, 
                V_{tq}^\ast \, V_{tb} \, \left\{ 
                {C_9}^{\rm eff} \, 
                        \left[ \bar{q} \, \gamma_\mu \, L \, b \right] \, 
                        \left[ \bar{\l} \, \gamma^\mu \, \l \right]
                + {C_{10}}^{\rm eff} \,  \left[ \bar{q} \, \gamma_\mu \, L \, b \right] \, 
                        \left[ \bar{\l} \, \gamma^\mu \, \gamma_5 \, \l \right]
                \right. \nonumber \\
        & &     \; \; \; \; \; \; \; \; \; \; \; \; \; \; \; \; \; \; \; \; \; \; \; \; \; \; \left.
                - 2 \, {C_7}^{\rm eff} \, 
                        \left[ \bar{q} \, i \, \sigma_{\mu \nu} \, 
                        \frac{\q^\nu}{\s} \left( R + \mq \, L \right) \, b
                        \right] 
                        \left[ \bar{\l} \, \gamma^\mu \, \l \right]
                \right\} \; .
                \label{eq:heff}
\end{eqnarray}
where $q^\mu$ denotes four-momentum of the dilepton and 
$s = q^2$. Notations with hat on the top means it is 
normalized with the $b-$quark mass. In the tree-level approximation, 
new interactions in Eq. (\ref{eq:z}) contributes to ${\cal O}_9$
and ${\cal O}_{10}$. 
Therefore, involving the continuum and resonances parts 
into calculation gives 
\begin{eqnarray}
	{C_7}^{\rm eff} & = & C_7 + 
		\eta_{\rm QCD}^{bq\gamma} \, {C_7}^{\rm new} \; ,
	\label{eq:c7eff} \\
        {C_9}^{\rm eff} & = & 
		\left( C_9 + {C_9}^{\rm new} \right) \left[ 
                1 + \frac{\alpha_s(\mu)}{\pi} \omega(\s) \right]
		+ {C_9}^{\rm con}(\s) + {C_9}^{\rm res}(\s) \; ,
        \label{eq:c9eff} \\
	{C_{10}}^{\rm eff} & = & C_{10} + {C_{10}}^{\rm new} \; , 
	\label{eq:c10eff}
\end{eqnarray}
where 
\begin{eqnarray}
        {C_9}^{\rm con}(\s) & = & 
                \left[ \left( 1 + 
                \frac{V_{uq}^\ast \, V_{ub}}{V_{tq}^\ast \, V_{tb}} -
		\frac{z_{qb}}{V_{tq}^\ast \, V_{tb}} \right) 
                g(\mc, \s) 
                - \frac{V_{uq}^\ast \, V_{ub}}{V_{tq}^\ast \, V_{tb}} 
                        g(\muq, \s) \right] 
        \nonumber \\
        & & \; \; \; \; \; \times \left(3 \, C_1 + C_2 + 3 \, C_3
                        + C_4 + 3 \, C_5 + C_6 \right)
                \nonumber \\
        & & - \frac{1}{2} g(1,\s) \left( 4 \, C_3 + 4 \, C_4 + 3 \,
                C_5 + C_6 \right) 
                \nonumber \\
        & & - \frac{1}{2} g(0,\s) \left( C_3 + 3 \, C_4 \right) 
                + \frac{2}{9} \left( 3 \, C_3 + C_4 +
                        3 \, C_5 + C_6 \right) \; , 
        \label{eq:c9con} \\
        {C_9}^{\rm res}(\s) & = & -\frac{16 \, \pi^2}{9} 
                \left( 3 \, C_1 + C_2 + 3 \, C_3 + C_4 + 
                        3 \, C_5 + C_6 \right)
        \nonumber \\    
        & &     \times \left[ 
                \left( 1 + 
                \frac{V_{uq}^\ast \, V_{ub}}{V_{tq}^\ast \, V_{tb}} - 
		\frac{z_{qb}}{V_{tq}^\ast \, V_{tb}} \right) 
                \sum_{V=\psi,\cdots} F_V(\s) 
                - \frac{V_{uq}^\ast \, V_{ub}}{V_{tq}^\ast \, V_{tb}}
                \sum_{V=\rho,\omega} F_V(\s) \right] \; , 
        \label{eq:c9res}\\
	{C_7}^{\rm new} & = & -\frac{1}{3} \, 
		\frac{z_{qb}}{V_{tq}^\ast \, V_{tb}} \; ,
	\label{eq:c7new}\\
	{C_9}^{\rm new} & = & 
		\frac{\pi}{\alpha} \, \frac{z_{qb}}{V_{tq}^\ast \, V_{tb}} 
		\left( 4 \, \sin^2 \theta_W - 1 \right) \; ,
	\label{eq:c9new}\\
	{C_{10}}^{\rm new} & = & 
		\frac{\pi}{\alpha} \, \frac{z_{qb}}{V_{tq}^\ast \, V_{tb}} \; .
	\label{eq:c10new}
\end{eqnarray}
Here $C_i$ ($i = 1, \cdots, 10$) are the Wilson coefficients
for each operator ${\cal O}_i$ calculated in the SM including 
the QCD corrections \cite{qcd}, $\omega(\s)$ represents the 
$O(\alpha_s)$ correction from the one gluon 
exchange in the matrix element of ${\cal O}_9$, 
$g(\hat{m}_{u_i}, \s)$ describes the continuum part
of $u_i\bar{u}_i$ pair contributions ($u_i = u, c$) and 
lastly $F_V(\s)$ denotes the resonances due to vector 
mesons including its momentum dependences. 
Remark that I keep ${C_7}^{\rm new}$, although it is occured
in the one-loop level, to get a constraint for the mixing ratio 
${z_{sb}}/{V_{ts}^\ast \, V_{tb}}$ from 
$B \rightarrow X_s \, \gamma$ decay which all contributions 
in the magnetic moment operator are coming from the one-loop level.
However in the discussion of 
$B \rightarrow X_q \, \l^+ \, \l^-$ one can ignore it. 
The result of ${C_7}^{\rm new}$ is under a natural assumption that 
$z_{\alpha \alpha} \sim 1$ and an approximation that 
$\left( {m_{d_i}}/{m_Z} \right)^2 \sim 0$ \cite{bqg}. 
Note that the dependence  
on the extra down-type quark mass ($m_{d_4}$) is supressed
due to the assumption.
I also assume the same QCD correction factor for both 
SM and new diagrams for all processes discussed in the present 
letter. This should be a good approximation since QCD 
corrections above the scale of $m_Z$ are negligible. 
For efficiency, the reader should refer \cite{all} and references 
therein for explicit expressions of each auxiliary functions above
which are not given here. 
I use the following values for further analysis \cite{qcd} \\
\begin{tabular}{lllll}
        $C_1 = -0.2404$, & 
        $C_2 = 1.1032$, &
        $C_3 = 0.0107$, &
        $C_4 = -0.0249$, &
        $C_5 = 0.0072$, \\
        $C_6 = -0.0302$, &
        $C_7 = -0.3109$, &
        $C_8 = -0.1478$, &
        $C_9 = 4.1990$, &
        $C_{10} = -4.5399$,   
\end{tabular}\\
and $\eta_{\rm QCD}^{bq\gamma} = 0.6745$ by putting 
$\Lambda_{\rm QCD}^{(5)} = 0.214$ (GeV) and the 
renormalization scale $\mu = 5$ (GeV).

The additional terms proportional to the mixing ratios 
${z_{qb}}/{V_{tq}^\ast \, V_{tb}}$
in Eqs. (\ref{eq:c9new}) and (\ref{eq:c10new}) are coming from the 
tree $Z-$exchange diagram respectively, while in Eqs. (\ref{eq:c9con}) and 
(\ref{eq:c9res}) are due to the unitarity violation relation
in Eq. (\ref{eq:vckm}) in the calculation of 
$b \rightarrow q \, u_i \, \bar{u}_i$ processes.
One may expect that these terms will contribute to the CP
violation in the channel together with the usual contribution in 
${C_9}^{\rm eff}$, because generally $z_{qb}$ has different phases with 
$V_{tq}^\ast \, V_{tb}$, i.e.
\begin{equation}
	\frac{z_{qb}}{V_{tq}^\ast \, V_{tb}} = 
	\left| \frac{z_{qb}}{V_{tq}^\ast \, V_{tb}} \right| \, 
	{\rm e}^{i \theta_q} \; ,
	\label{eq:mr} 
\end{equation} 
where $\theta_q = {\rm arg}\left( {z_{qb}}/{V_{tq}^\ast \, V_{tb}}\right)$. 
However, as pointed out later ${C_{10}}^{\rm new}$ contribute 
nothing to the CP asymmetry, while ${C_9}^{\rm new}$ gives 
small change. 
From Eqs. (\ref{eq:c9new}) and (\ref{eq:c10new}), one can predict 
easily that ${C_i}^{\rm new}$ ($i : 9, 10$) should be large even for 
small mixing ratios, i.e. 
$\left| {z_{qb}}/{V_{tq}^\ast \, V_{tb}} \right| \sim O({10}^{-2})$. 
Especially, a large enhancement is expected in ${C_{10}}^{\rm eff}$, 
because of no suppression due to Weinberg angle. High dependences 
of ${C_{10}}^{\rm eff}$ are expected in the forward-backward (FB) 
and lepton-polarization (LP) asymmetries.

Before going on analysing the decay rates and asymmetries, 
I consider experimentally well-known $B_d^0 - \bar{B}_d^0$ 
mixing and $B \rightarrow X_s \, \gamma$ to obtain some constraints 
for the mixing ratios. First, from the $B_d^0 - \bar{B}_d^0$ mixing,
a constraint for ${z_{db}}/{V_{td}^\ast \, V_{tb}}$ can be obtained 
from the measurement of $x_d$. 
In general $B_q^0 - \bar{B}_q^0$ mixings, $x_q$ is given as \cite{bb}
\begin{equation}
	x_q = \c_{B_q\bar{B}_q} \, \left| F_{\Delta B=2} \right| \, 
		\left| V_{tq}^\ast \, V_{tb} \right|^2 \, 
		\left( 1 + \frac{4 \pi \, \sin^2 \theta_W}{
		\alpha \, \left| F_{\Delta B=2} \right|} 
		\left| \frac{z_{qb}}{V_{tq}^\ast \, V_{tb}} \right|^2 \, 
		{\rm e}^{2 i \theta_q} \right) \; ,
	\label{eq:xq}
\end{equation}
where $\c_{B_q\bar{B}_q} = {{G_F}^2}/{(6 \pi^2)} \, \tau_{B_q} \, 
\eta_{\rm QCD}^{B\bar{B}} \, m_{B_q} \, {m_W}^2 \, \left( {f_{B_q}}^2 \, B_q \right)$. 
Numerically, using QCD correction factor 
$\eta_{\rm QCD}^{B\bar{B}} = 0.55$, $m_W = 80.33$ (GeV) and  
$\sqrt{{f_{B_d}}^2 \, B_{B_d}} = 173 \pm 40$ (MeV), 
one obtains $\c_{B_d\bar{B}_d} = 13834.6^{+7137.1}_{-5657.9}$. 
Here, for $m_{B_d}$ and $\tau_{B_d}$, I use $m_{B^0} = 5279.2 \pm 1.8$ (MeV) 
and $\tau_{B^0} = 1.28 \pm 0.06$ (ps) and from the box diagram calculation 
in the SM, $\left| F_{\Delta B=2} \right| = 0.543$
for $m_t = 175$ (GeV). On the other 
hand, recent experiment gives $x_d = 0.73 \pm 0.05$ \cite{pdg}.
Secondly, from $B \rightarrow X_s \, \gamma$ which experimentally 
has been measured to be ${\cal B}(B \rightarrow X_s \, \gamma) = 
\left( 2.32 \pm 0.57 \pm 0.35 \right) \times {10}^{-4}$ \cite{cleo}, 
one can extract a constraint for ${z_{sb}}/{V_{ts}^\ast \, V_{tb}}$. 
Generally in terms of the semi-leptonic $B$ decay, the branching 
ratio for $B \rightarrow X_q \, \gamma$ is expressed as
\begin{equation}
	{\cal B}(B \rightarrow X_q \, \gamma)	= 
		\c_{bq\gamma} \, 
		\left| V_{tq}^\ast \, V_{tb} \right|^2 \, 
		\left| {C_7}^{\rm eff} \right|^2 \; , 
	\label{eq:bqg}
\end{equation}
where $\c_{bq\gamma} = 
{\cal B} (B \rightarrow X_c \, \l \, \bar{\nu}) \, 
{\left(6 \, \alpha \right)}/{
\left(\pi \, f(\mc) \, \kappa(\mc) \, \left| V_{cb} \right|^2\right)}$. 
The value is $\c_{bq\gamma} = {1.910}^{+0.399}_{-0.315}$ by 
using $\left| V_{cb} \right| = 0.041 \pm 0.003$, 
${\cal B} (B \rightarrow X_c \, \l \, \bar{\nu}) = (10.4 \pm 0.4)\%$, 
while $f(\mc) = 0.542$ and $\kappa(\mc) = 0.885$ for 
$\mc = 0.29$ \cite{all}.

Lastly, substituting the experiment results of $x_d$ and 
$B \rightarrow X_s \, \gamma$ into Eqs. (\ref{eq:xq}) and (\ref{eq:bqg}),  
one obtains the bounds for the mixing ratios as depicted in 
Fig. 1. In the left figure, the solid and dashed 
curves denote the central, upper and lower bounds for $\left| z_{db} \right|$
as a function of $\left| V_{td}^\ast \, V_{tb} \right|$, while  
in the right one the bounds are given for $q = s$ with
varying $\theta_s$.
Since my emphasis is on a case that a relatively small CKM unitarity 
violation may change the distributions in the decays, further I put both 
mixing ratios to be small, that is  
$\left| {z_{qb}}/{V_{tq}^\ast \, V_{tb}} \right| \sim 0.01$
which correspond to $\left| V_{td}^\ast \, V_{tb} \right| \sim 0.0095$ and 
$\left| V_{ts}^\ast \, V_{tb} \right| \sim 0.036$. 
Because of the smallness, its contributions in Eqs. (\ref{eq:c7eff}), 
(\ref{eq:c9con}) and (\ref{eq:c9res}) can be ignored, while 
in Eqs. (\ref{eq:c9new}) and (\ref{eq:c10new}) must be kept
as it stands. Particularly in the case of $q = s$ the ratio 
${V_{uq}^\ast \, V_{ub}}/{V_{tq}^\ast \, V_{tb}}$ 
is negligible, while for $q = d$ it should be kept and roughly 
I put the ratio and the phase to be same with the SM \cite{all}
for simplicity. 

Now I turn to analyse the dilepton invariant mass distribution 
of the decay rates and asymmetries in the channels. Involving 
the lepton and quark masses, the differential branching ratio (BR) is 
\begin{eqnarray}
        \frac{{\rm d}{\cal B}}{{\rm d}\s} & = & 
      	\int_{-1}^1 {\rm d}z \, \frac{{\rm d}^2{\cal B}}{{\rm d}\s {\rm d}z}
	\nonumber \\	
        & = & \frac{4}{3} \, \c_{bq\l\l} \, 
	\sqrt{1 - \frac{4 \, \ml^2}{\s}} \, \u(\s) \left\{ 
        6 \, \left[ \left| {C_9}^{\rm eff} \right|^2 - \left| {C_{10}}^{\rm eff} \right|^2 \right] 
                \, \ml^2 \, \left[1 - \s + \mq^2 \right]
        \right. 
        \nonumber \\
        & & \left. 
        + \left[ \left| {C_9}^{\rm eff} \right|^2 + \left| {C_{10}}^{\rm eff} \right|^2 \right]
                \left[ (1 - \mq^2)^2 + \s \, (1 + \mq^2) - 
                2 \, \s^2 + {\u(\s)}^2 \frac{2 \, \ml^2}{\s} \right]
        \right. 
        \nonumber \\
        & & \left. 
        + 4 \, \left| {C_7}^{\rm eff} \right|^2 \frac{1 + {2\, \ml^2}/\s}{\s}
        \right. 
        \nonumber \\
        & & \left. \; \; \; \;
                \times \left[ 2 \, (1 + \mq^2)(1 - \mq^2)^2 - 
                (1 + 14 \, \mq^2 + \mq^4) \, \s - (1 + \mq^2) \, \s^2 \right]
        \right. 
        \nonumber \\
        & & \left. 
        + 12 \, {\rm Re} \left( {C_9}^{\rm eff} \right)^\ast \, {C_7}^{\rm eff} \,
                \left[ 1 + \frac{2 \, \ml^2}{\s} \right] \, 
                \left[ (1 - \mq^2)^2 - (1 + \mq^2) \, \s \right]
        \right\} \; .
        \label{eq:dbr}
\end{eqnarray}
where $\u(\s) = \sqrt{[\s - (1 + \mq)^2][\s - (1 - \mq)^2]}$,
$z = \cos \theta$ is the angle of $\l^+$ measured with 
respect to the $b-$quark direction in the dilepton CM system and 
$\c_{bq\l\l} = {\cal B} (B \rightarrow X_c \, \l \, \bar{\nu}) 
{\left( 3 \, \alpha^2 \, \left| V_{tq}^\ast \, V_{tb} \right|^2 \right)}/{
\left(16 \, \pi^2 \, \left| V_{cb} \right|^2 \, f(\mc) \, \kappa(\mc) \right)}$. 

The normalized forward-backward (FB) asymmetry is defined \cite{fba} 
and calculated as follows
\begin{eqnarray}
        \bar{\cal A}_{\rm FB} & = & \frac{\displaystyle 
        \int_0^1 {\rm d}z \, 
        \frac{{\rm d}^2{\cal B}}{{\rm d}\s {\rm d}z}
        - \int_{-1}^0 {\rm d}z \, 
        \frac{{\rm d}^2{\cal B}}{{\rm d}\s {\rm d}z}
        }{\displaystyle 
        \int_0^1 {\rm d}z \, 
        \frac{{\rm d}^2{\cal B}}{{\rm d}\s {\rm d}z}
        + \int_{-1}^0 {\rm d}z \, 
        \frac{{\rm d}^2{\cal B}}{{\rm d}\s {\rm d}z}
        } 
	\nonumber \\
        & = & \frac{-4 \, \c_{bq\l\l}}{ 
		{{\rm d}{\cal B}(\s)}/{{\rm d}\s}}
        \sqrt{1 - \frac{4 \, \ml^2}{\s}} \, {\u(\s)}^2 \, 
	C_{10} \, \left[  
        {\rm Re} \, \left( {C_9}^{\rm eff} \right)^\ast \, \s 
        + 2 \, {C_7}^{\rm eff} \, \, \left( 1 + \mq^2 \right)
        \right] \; . 
        \label{eq:dfba}
\end{eqnarray}

Doing same treatment as \cite{cpvf} in the 
amplitude level, the normalized CP asymmetry can be 
written simply as
\begin{equation}
        \bar{\cal A}_{\rm CP} = \frac{\displaystyle
        {{\rm d}{\cal B}}/{{\rm d}\s}
        - 
        {{\rm d}\bar{\cal B}}/{{\rm d}\s}
        }{\displaystyle 
        {{\rm d}{\cal B}}/{{\rm d}\s}
        + 
        {{\rm d}\bar{\cal B}}/{{\rm d}\s}
        }       
        = \frac{\displaystyle
        -2 \, {{\rm d}{\cal A}_{\rm CP}}/{{\rm d}\s}
        }{
        {{\rm d}{\cal B}}/{{\rm d}\s} + 
        2 \, {{\rm d}{\cal A}_{\rm CP}}/{{\rm d}\s}} \; ,
\end{equation}
where $\cal B$ and $\bar{\cal B}$ denote the BR of the 
$\bar{b} \rightarrow q \, \l^+ \, \l^-$ and its complex conjugate 
$b \rightarrow \bar{q} \, \l^+ \, \l^-$ respectively.
Different with the SM, in the present model there is a 
possible new source of CP violation in ${C_{10}}^{\rm eff}$. 
Redefine the Wilson coefficients ${C_i}^{\rm eff}$ ($i : 9, 10$)
as ${C_i}^{\rm eff} = \bar{C}_i + 
\left( {V_{uq}^\ast \, V_{ub}}/{V_{tq}^\ast \, V_{tb}} \right) \, {C_i}^{\rm CP} + 
\left( {z_{qb}}/{V_{tq}^\ast \, V_{tb}} \right) \, {\bar{C}_i}^{\rm CP}$, 
the result is
\begin{eqnarray}
 	\frac{{\rm d}{\cal A}_{\rm CP}}{{\rm d}\s} & = & 
	\frac{4}{3} \c_{bq\l\l} \, 
       	\sqrt{1 - \frac{4 \,\ml^2}{\s}} \, \u(\s) \, 
	\left\{ 
	6 \left[ \c_9 - \c_{10} \right] \, \ml^2 \, 
		\left[1 - \s + \mq^2 \right] \right. 
	\nonumber \\
	& & \left. 
	+ \left[ \c_9 + \c_{10} \right] \, 
		\left[ (1 - \mq^2)^2 + \s \, (1 + \mq^2) - 
		2 \, \s^2 + {\u(\s)}^2 \frac{2 \, \ml^2}{\s}
	\right] \right. 
	\nonumber \\
	& & \left. 
	+ 6 \, \bar{\c}_9 \, {C_7}^{\rm eff} \, \left[ 1 + \frac{2 \, \ml^2}{\s} \right] \, 
		\left[ (1 - \mq^2)^2 - (1 + \mq^2) \, \s \right]
	\right\} \; ,
	\label{eq:dcpa}
\end{eqnarray}
where
\begin{eqnarray}
	\c_i & = &
	{\rm Im} \left( \frac{V_{uq}^\ast \, V_{ub}}{V_{tq}^\ast \, V_{tb}} \right) \, 
		{\rm Im} \left( {\bar{C}_i}^\ast \, {C_i}^{\rm CP} \right) + 
	{\rm Im} \left( \frac{z_{qb}}{V_{tq}^\ast \, V_{tb}} \right) \, 	
		{\rm Im} \left( {\bar{C}_i}^\ast \, {\bar{C}_i}^{\rm CP} \right) 
	\nonumber \\
	& & + 
	{\rm Im} \left( \left( \frac{V_{uq}^\ast \, V_{ub}}{V_{tq}^\ast \, V_{tb}} \right)^\ast
		\frac{z_{qb}}{V_{tq}^\ast \, V_{tb}} \right) \, 
		{\rm Im} \left( {{C_i}^{\rm CP}}^\ast \, {\bar{C}_i}^{\rm CP} \right) \; , 
	\label{eq:fi} \\
	\bar{\c}_i & = & 
	{\rm Im} \left( \frac{V_{uq}^\ast \, V_{ub}}{V_{tq}^\ast \, V_{tb}} \right) \, 
		{\rm Im} \left( {C_i}^{\rm CP} \right) + 
	{\rm Im} \left( \frac{z_{qb}}{V_{tq}^\ast \, V_{tb}} \right) \, 
		{\rm Im} \left( {\bar{C}_i}^{\rm CP} \right) \; .
	\label{eq:fibar}
\end{eqnarray}
As mentioned before, it is clear that ${C_{10}}^{\rm eff}$ will not contribute 
to the CP asymmetry in the model. Because one needs both complex 
CKM factor and complex Wilson coefficients to produce CP asymmetry,
that is ${\rm Im} \left( \bar{C}_i \right)$, 
${\rm Im} \left( {C_i}^{\rm CP} \right)$, 
${\rm Im} \left( {\bar{C}_i}^{\rm CP} \right)$ and 
the imaginary of its combination must not be zero.

For the longitudinal polarization in the $\l^-$ rest frame, the normalized 
lepton-polarization (LP) asymmetry is given as \cite{lpa}
\begin{eqnarray}
        \bar{\cal A}_{\rm LP} & = & 
        \frac{\displaystyle {{\rm d}{\cal B}(\nb_{\ell^-})}/{{\rm d}\s}
        - 
        {{\rm d}{\cal B}(-\nb_{\ell^-})}/{{\rm d}\s}
        }{\displaystyle {{\rm d}{\cal B}(\nb_{\ell^-})}/{{\rm d}\s}
        + 
        {{\rm d}{\cal B}(-\nb_{\ell^-})}/{{\rm d}\s}
        }       
	\nonumber \\
        & = & \frac{8 \, \c_{bq\l\l}}{
		3 \, {{\rm d}{\cal B}(\s)}/{{\rm d}\s}}
        \left(1 - \frac{4 \,\ml^2}{\s} \right) \, \u(\s) \, 
	{C_{10}}^{\rm eff} \, \left\{ 6 \, {C_7}^{\rm eff} \, \left[ (1 - \mq^2)^2 - 
                \s \, (1 + \mq^2) \right] 
        \right.
        \nonumber \\
        & & \left. \; \; \; \; \; \; 
        + {\rm Re} \left( {C_9}^{\rm eff} \right)^\ast \, 
        \left[ (1 - \mq^2)^2 + \s \, (1 + mq^2) - 2 \, \s^2 \right] 
        \right\} \; .
        \label{eq:dlpa}
\end{eqnarray}
with $\nb_{\ell^-} = {\pb_{\ell^-}}/{|\pb_{\ell^-}|}$. 

As the results, the distributions of differential BR and 
asymmetries on dilepton invariant mass are plotted in Figs. 2$\sim$4.
New contribution in the model changes the differential BR
significantly for both $q = d, s$. It also contributes 
to FB and LP asymmetries without no significant differences 
for both $q = d, s$ since the overwhole CKM factor is 
eliminated by definition. Furthermore, in the differential BR the 
distribution is dominated by ${C_9}^{\rm eff}$, while in 
the FB and LP asymmetries are dominated by ${C_{10}}^{\rm eff}$. 
On the other hand, the new source of CP violation does not 
contribute to the CP asymmetry as large as expected.
However, for $q = s$ the new term 
in ${C_9}^{\rm eff}$ enhance the CP asymmetry for about 
one order. Note that in the SM, $\bar{\cal A}_{\rm CP} 
(B \rightarrow X_s \, e^+ \, e^-) \sim O({10}^{-3})$.

In conclusion, the measurements of the decay rates and 
asymmetries in $B \rightarrow X_q \, \l^+ \, \l^-$  
decays together with the measurement of CP asymmetry 
in flavor changing charge-curent $B$ decays will provide 
a crucial test of CKM unitarity as well as leading to the 
discovery of unitarity violation.

I would like to thank the Ministry of Education, Science and
Culture (Monbusho - Japan) for financial support during my 
stay in Japan.

\clearpage

\begin{figure}[t]
        \begin{minipage}[t]{\minitwocolumn}
        \begin{center}
% GNUPLOT: LaTeX picture using EEPIC macros
\setlength{\unitlength}{0.240900pt}

        \end{center}
        \end{minipage}
        \caption{Allowed values for $\left| z_{db} \right|$ (left) and 
	$\left| z_{sb} \right|$ (right) according
	to its counterpart $\left| V_{td}^\ast \, V_{tb} \right|$ and 
	$\left| V_{ts}^\ast \, V_{tb} \right|$.}
        \label{fig:bound}
\end{figure}    

\begin{figure}[h]
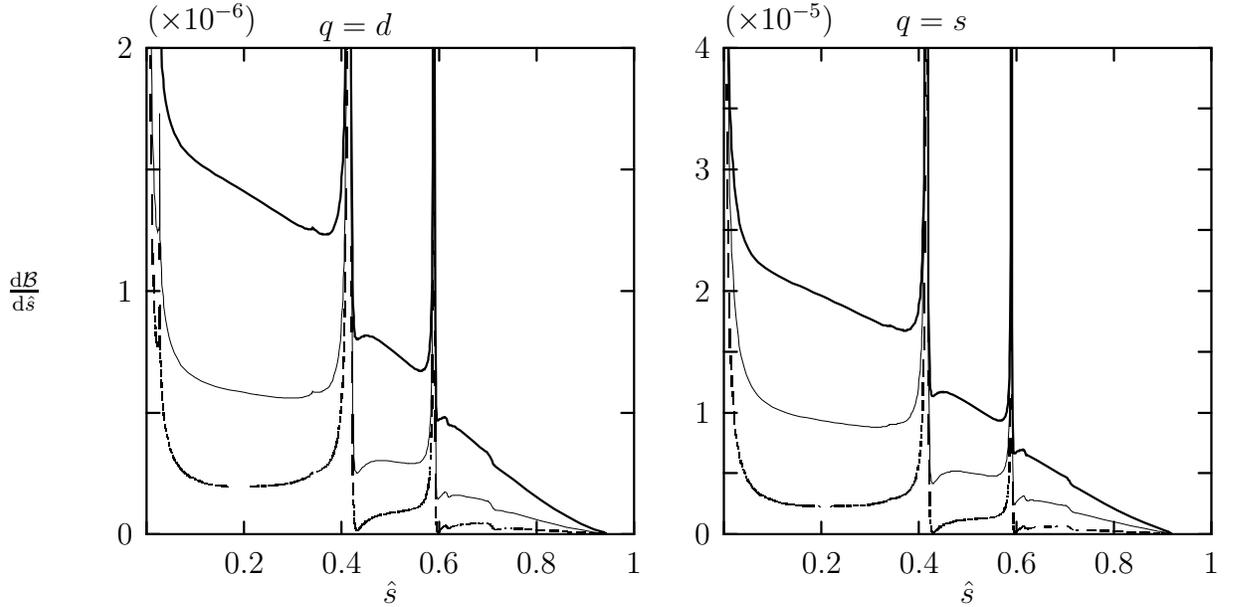

        \begin{minipage}[t]{\minitwocolumn}
        \begin{center}
% GNUPLOT: LaTeX picture using EEPIC macros
\setlength{\unitlength}{0.240900pt}
%
        \end{center}
        \end{minipage}
        \caption{Differential BR for $e^+ e-$ in the SM (thin solid curve) 
	and in the present model with $\theta_q = 0^o$ (dashed curve)
	and $\theta_q = {180}^o$ (solid thick curve).}
        \label{fig:dbr}
\end{figure}   

\clearpage

\begin{figure}[t]
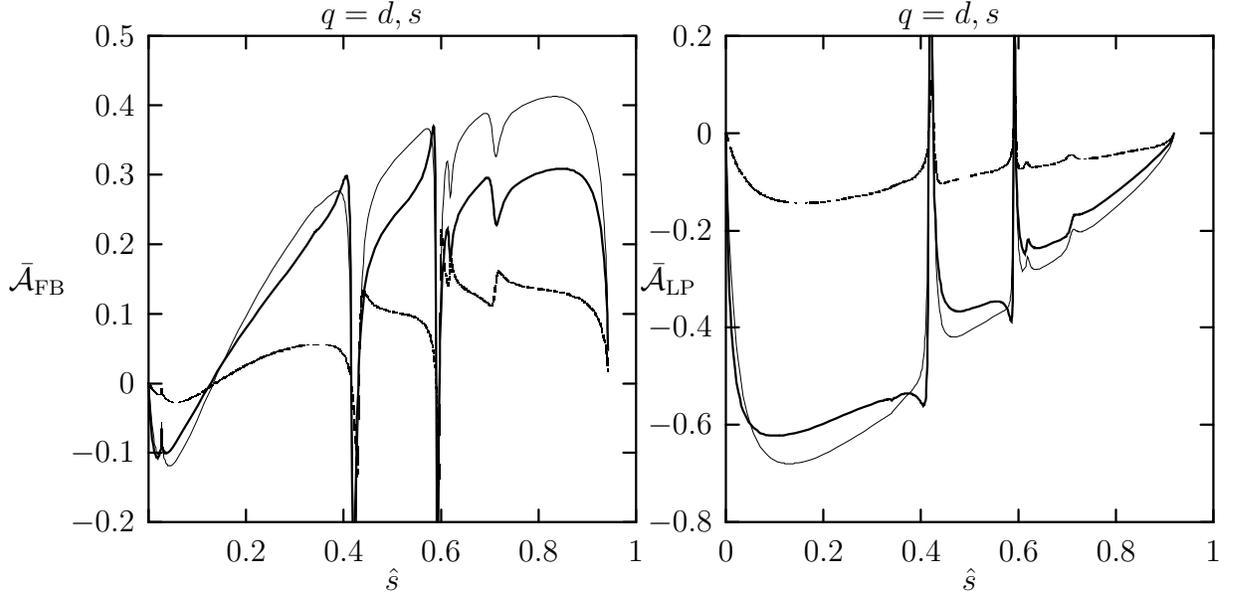

        \begin{minipage}[t]{\minitwocolumn}
        \begin{center}
% GNUPLOT: LaTeX picture using EEPIC macros
\setlength{\unitlength}{0.240900pt}
%
	\end{center}
        \end{minipage}
        \caption{FB (left) asymmetry for $e^+ e^-$ and 
	LP asymmetry for $\mu^+ \mu^-$ in the SM (thin solid curve) 
	and in the present model with $\theta_q = 0^o$ (dashed curve)
	and $\theta_q = {180}^o$ (solid thick curve).}
        \label{fig:dfba}
\end{figure}   

\begin{figure}[h]
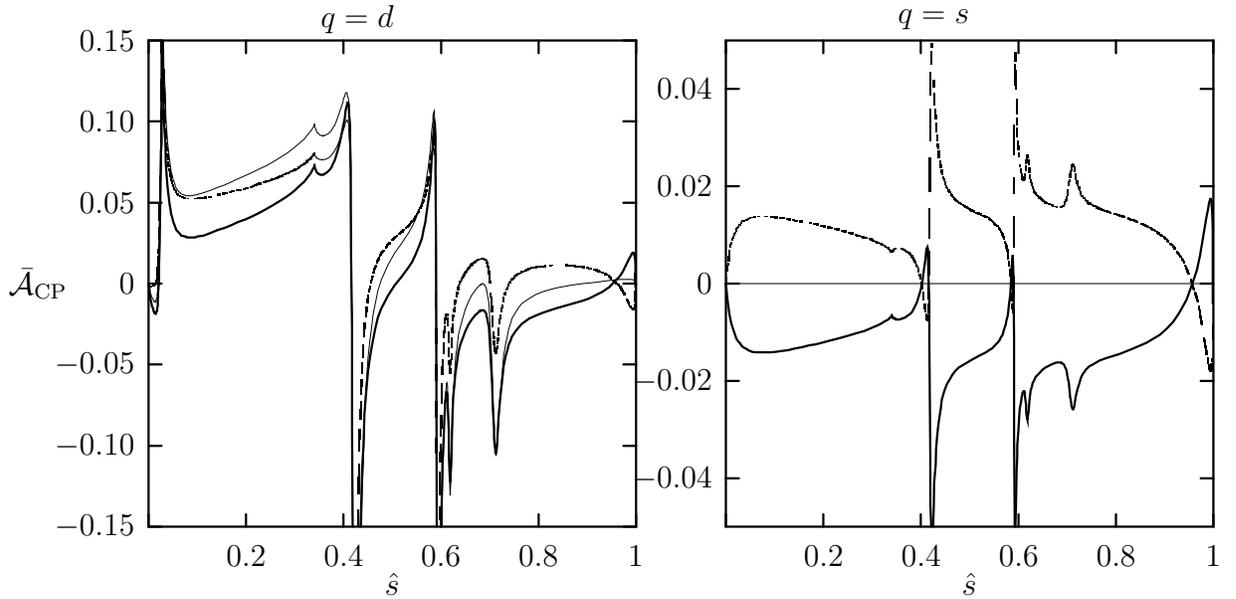

        \begin{minipage}[t]{\minitwocolumn}
        \begin{center}
% GNUPLOT: LaTeX picture using EEPIC macros
\setlength{\unitlength}{0.240900pt}
%
        \end{center}
        \end{minipage}
        \caption{CP asymmetry for $e^+ e^-$ in the SM (thin solid curve) 
	and in the present model with $\theta_q = {90}^o$ (solid thick curve)
	and $\theta_q = {-90}^o$ (dashed curve).}
        \label{fig:dcpa}
\end{figure}   

\end{document}